\documentclass[aps,prl,twocolumn,groupedaddress]{revtex4}
\usepackage{graphicx}

\begin{document}

\title{Regularization of moving boundaries in a Laplacian field\\
by a mixed Dirichlet-Neumann boundary condition --- exact results}

\author{Bernard Meulenbroek$^1$, Ute Ebert$^{1,2}$, Lothar Sch\"afer$^3$}

\affiliation{$^1$CWI, P.O.Box 94079, 1090 GB Amsterdam, The Netherlands,}
\affiliation{$^2$Dept. Physics, Eindhoven Univ. Techn., The Netherlands,}
\affiliation{$^3$Dept. Physics, Univ. Essen, Germany.}

\date{\today}

\newcommand \be{\begin{equation}}
\newcommand \ee{\end{equation}}
\newcommand \ba{\begin{eqnarray}}
\newcommand \ea{\end{eqnarray}}
\def\nn{\nonumber}
\def\np{\newpage}

\begin{abstract} 
The dynamics of ionization fronts that generate
a conducting body, are in simplest approximation 
equivalent to viscous fingering without regularization. 
Going beyond this approximation, we suggest that ionization fronts 
can be modeled by a mixed Dirichlet-Neumann boundary condition. 
We derive exact uniformly propagating solutions of this problem in 2D
and construct a single partial differential equation governing 
small perturbations of these 
solutions. For some parameter value, this equation can be solved 
analytically which shows that the uniformly propagating solution 
is linearly convectively stable.
\end{abstract}
\pacs{INSERT PACS}
\maketitle

Boundaries between two phases that move according to the gradient
of a Laplacian or diffusive field, occur in many fields of the
natural sciences and have a long and intricate research history 
\cite{Pelce}; well known examples 
include viscous fingering in Hele-Shaw flow \cite{st,visfing}, 
solidification fronts in under-cooled melts \cite{Pelce},
migration of steps \cite{BCF} or electromigration of voids
\cite{electro0, electro} on the surface of layered solids 
or boundaries of bacterial colonies 
in an external nutrition field \cite{bact}. 
Viscous fingering here takes a paradigmatic role as the oldest
and most studied problem --- determining the long time
dynamics up to today leads to mathematical surprises 
\cite{Tanveer,Tanveer2,KL,Jaume}.

A similar moving boundary problem arises in 
so-called streamer discharges \cite{PREUWC,PRLMan}
that precede sparks and lightning.
Streamer ionization fronts can be understood as moving boundaries 
separating an ionized phase from a non-conducting phase
\cite{PRLMan,Man04,Bern1}. The inner front structure can
be approximated by a
boundary condition of mixed Dirichlet-Neumann-type, as we will
sketch below. A similar boundary condition appears in step motion
on the surface of layered solids when the Schwoebel
barrier is taken into account \cite{BCF}.
Our boundary condition has a similar physical effect as the 
curvature correction in viscous fingering. We do show here 
that it indeed stabilizes certain uniformly translating shapes.
In our analysis below, we encountered a number of surprises: 
$(i)$ planar fronts are linearly unstable to transversal perturbations 
of arbitrary wave vector $0<k<\infty$, still we find that sufficiently 
curved fronts are linearly convectively stable; 
$(ii)$ a simple explicit uniformly 
translating solution can always be found; $(iii)$ 
linear perturbations of these solutions can be reformulated
in terms of a single partial differential equation,
$(iv)$ if the solution of this perturbation
problem is Taylor expanded and truncated at any finite order, 
the eigensolutions seem to be purely oscillating, $(v)$ however, 
for a particular parameter value, the linear perturbation theory 
has an explicit analytical solution that shows that there 
are no oscillations. Rather, perturbations might grow 
for some time, while they simultaneously are convected to the back
where they disappear. Only a shift of the original shape remains for 
$t\to\infty$. To our knowledge, this is the first explicit solution
showing the convective stabilization of a curved front according
to the concept of Zeldovich \cite{comb}.

In fact, the interfacial dynamics with our boundary condition 
can be addressed by explicit analysis much further than the classical 
viscous fingering problem.
It therefore might contribute not only to the understanding
of ionization fronts, but also shed new light on
other moving boundary problems like the classical viscous fingering problem.


A simple moving boundary approximation for a streamer ionization front 
was suggested by Lozansky and Firsov \cite{Firsov}: The front penetrates 
into a non-ionized and electrically neutral region (indicated with 
a ${}^+$) with a velocity determined by the local electric field
${\bf E}^+=-\nabla\varphi^+$:
\be
\label{phi+}
\nabla^2\varphi^+=0~,~~~
\label{vn}
v_n = \hat{\bf n}\cdot \nabla\varphi^+,
\ee
where $\hat{\bf n}$ is the local normal on the boundary.
Approximating the interior ionized region as ideally conducting
\be
\label{phi-}
\varphi^-={\rm const.},
\ee
and the electric potential as continuous across the ionization
boundary $\varphi^+=\varphi^-$, one arrives at the Lozansky-Firsov 
interfacial model. This model was suggested in \cite{PRLMan} to explain
streamer branching, and it was explicitly analyzed in \cite{Bern1}.
Replacing the electric potential $\varphi$ by the pressure field $p$,
one finds the non-regularized motion of viscous fingers in 
a Hele-Shaw cell. The model generically leads within finite time
to the formation of cusps, i.e., of locations on the interface 
with vanishing radius of curvature \cite{entov}. 


We here propose to replace the boundary condition $\varphi^+-\varphi^-=0$ by 
\be
\label{bcn}
\varphi^+-\varphi^-=\epsilon \;\hat{\bf n}\cdot \nabla\varphi^+
\ee
to suppress these unphysical cusps.
Here the length scale $\epsilon$ characterizes the width of
the ionization front where the ionization increases and 
the electric field decreases. It determines the jump 
$\varphi^+-\varphi^-$ of the electric potential across 
the boundary for given field $\nabla\varphi^+$ ahead of the front.

The classical boundary condition for viscous fingering is
$\varphi^+-\varphi^-=\gamma\kappa$ where $\kappa$ 
is the local curvature of the moving interface, 
and $\gamma$ is surface tension.
In contrast, the boundary condition (\ref{bcn})
does not involve front curvature, but 
can be derived from {\em planar} ionization fronts, more precisely
from a minimal set
of partial differential equations for electron and ion densities and
their coupling to the electric field \cite{PREUWC}.
The formal derivation will be given elsewhere. Here we note that 
ignoring electron diffusion ($D=0$) as in \cite{Man04}, the planar
uniformly translating front solutions of the p.d.e.s always yield
a relation $\varphi^+-\varphi^-=F(\hat{\bf n}\cdot\nabla\varphi^+)$.
For large field $E^+=\hat{\bf n}\cdot\nabla\varphi^+$ ahead of the front,
the function $F$ becomes linear, 
and the boundary condition (\ref{bcn}) results.

This boundary condition has a similar physical
effect as the curvature correction in viscous fingering: 
high local fields ahead of the front decrease due to
the change of $\varphi^+$ on the boundary, and the interface
moves slower than an equipotential interface (where $\varphi^+={\rm const}$).
While the boundary condition of viscous fingering
suppresses high interfacial curvatures that can lead 
to high fields, the boundary condition (\ref{bcn}) 
suppresses high fields that frequently are
due to high local curvatures. This physical consideration 
has motivated our present study whether the boundary condition 
(\ref{bcn}) also regularizes the interfacial motion.

The minimal p.d.e.\ model for streamer fronts with $D=0$
leads to a dispersion relation with asymptotes
\be
\label{disp}
s(k)=
\left\{\begin{array}{ll}vk &~~~\mbox{for }k\ll 1/\epsilon\\
v/\epsilon &~~~\mbox{for }k\gg 1/\epsilon\end{array}\right.
\ee
for linear transversal perturbations 
$e^{ikx+st}$ of planar interfaces 
\cite{Man04}. It is important to check whether 
the moving boundary approximation (\ref{phi+})--(\ref{bcn}) 
reproduces this behavior. Indeed, analyzing planar interfaces we
find
$s(k)=vk/(1+\epsilon k)$
in full agreement with (\ref{disp})
as we will show in detail elsewhere.
This strongly suggests that the interfacial model captures
the correct physics. It shows furthermore, that planar fronts
are linearly unstable against any wave-vector $k$ for all $\epsilon$. 

We now restrict the analysis to the two-dimensional version of the model
and to arbitrary closed streamer shapes 
in an electric field that becomes homogeneous 
\be
\label{asymp}
\varphi(x,y)\to -E_0 x~~~\mbox{far from the ionized body}.
\ee
The problem is treated with conformal mapping methods \cite{Bern1}:
The exterior of the streamer where
$\nabla^2\varphi^+=0$ can be mapped onto the interior of the unit circle.
Parameterizing the original space with $z=x+iy$ and the interior
of the unit disk with $\omega$, the position of the streamer
can be written as
\be
\label{ft}
z=x+iy=f_t(\omega)=\frac1{h_t(\omega)}=\sum_{k=-1}^\infty a_k(t)\;\omega^k,
\ee
where $h_t(\omega)$ is analytical on the unit disk with a single zero
at $\omega=0$ and therefore has the Laurent expansion
given on the right. The boundary of the ionized body 
\be
\omega=e^{i\alpha},~~~\alpha\in[0,2\pi[,
\ee
is parametrized by the angle $\alpha$.


The potential $\varphi^+$ is a harmonic function due to (\ref{phi+}),
therefore one can find a complex potential $\Phi(z)=\varphi^++i\psi$
that is analytic. Its asymptote is $\Phi(z)\to -E_0z$ for $|z|\to\infty$
according to (\ref{asymp}). For the complex potential $\hat\Phi(\omega)$,
this means that
\be
\label{hatPhi} \label{abc2}
\hat\Phi(\omega)=\Phi(f_t(\omega))=-E_0\;a_{-1}(t)\left(\frac1\omega
+\sum_{k=0}^\infty c_k(t)\;\omega^k\right),
\ee
where the pole $\propto 1/\omega$ stems from the constant far field
$E_0$, and the remainder is a Taylor expansion that accounts for 
the analyticity of $\hat\Phi$. 
The boundary motion $v_n=\hat{\bf n}\cdot\nabla\varphi$ 
(\ref{vn}) is rewritten as 
\be \label{eqdyn}
{\rm Re} \left[ i\partial _\alpha f_t^*\; \partial_t f_t \right]
={\rm Re} \left[-i\partial_\alpha \hat \Phi(e^{i\alpha})\right].
\ee
The boundary condition (\ref{bcn}) takes the form
\be \label{abc1}
{\rm Re} \left[\hat \Phi(e^{i\alpha})\right]= \epsilon\;{\rm Re}
\left[\frac{i\partial _\alpha \hat \Phi(e^{i\alpha})}
{|\partial_\alpha f_t|}\right].
\ee 
The moving boundary problem is now reformulated as Eqs.\ (\ref{eqdyn}) 
and (\ref{abc1}) together with the ans\"atze (\ref{ft}) and (\ref{abc2}) 
for $f_t$ and $\hat\Phi$.

For the unregularized problem (where $\epsilon=0$), it is well known that all 
ellipses with a main axis oriented parallel to the 
external field are uniformly translating solutions: for principal
radii $r_{x,y}=a_{-1}\pm a_1$, they propagate with velocity 
$v=-E_0(r_x+r_y)/r_y$, while the potential is
$\hat\Phi=E_0a_{-1}(t)(\omega-1/\omega)$ \cite{Bern1,entov}.

For a moving boundary problem with regularization, 
there are only rare cases where analytical solutions can be given,
and frequently they are given only implicitly \cite{SolVF,SolVF2,entov}. 
For the present problem, however, an explicit solution
is found for all $\epsilon>0$:
\ba
\label{Cir1}
z&=&f_t(\omega)=\frac{a_{-1}}{\omega}+vt,~~~\partial_ta_{-1}=0,
\\
\label{Cir2}
\hat\Phi(\omega)&=&-E_0a_{-1}\left(\frac1{\omega}
-\frac{1-\epsilon/a_{-1}}{1+\epsilon/a_{-1}}\;\omega\right).
\ea
This solution simply describes a circle 
$z=x+iy=a_{-1}e^{-i\alpha}+vt$ with radius $a_{-1}$ 
that according to (\ref{eqdyn}) propagates with velocity
$v=-2 E_0/(1+\epsilon/a_{-1})$.
Note that $\epsilon$ changes the velocity, but not the shape of
the solution. 
Note further that the multiplicity of uniformly translating 
solutions reduces through regularization in a similar way as in
viscous fingering, namely from a family of ellipse solutions
characterized by two continuous parameters $a_{-1}$ and $a_1$
to a family of circle solutions characterized by only one radius 
$a_{-1}$ or interface width $\epsilon$. 

The physical problem has two length scales, the interface width $\epsilon$
and the circle radius $a_{-1}$. In the sequel, we set $a_{-1}=1$,
measuring all lengths relative to the radius of the circle.

Now the question arises whether a uniformly translating circle
is stable against small perturbations, in particular, in view
of the linear instability of the planar front (\ref{disp}). 
The basic equations (\ref{ft})--(\ref{abc1}) show a quite complicated
structure, and it is a remarkable feature that
linear stability analysis of the translating circle 
(\ref{Cir1})--(\ref{Cir2}) can be reduced to solving a single 
partial differential equation. We write
\ba
f_t(\omega)&=&\frac1{\omega}+\tau+\beta(\omega,\tau),
~~~\tau=vt,~~~v=\frac{-2E_0}{1+\epsilon},\\
\hat\Phi(\omega)&=&-E_0\left(\frac1{\omega}-
\frac{1-\epsilon}{1+\epsilon}\;\omega\right)+v\;\phi(\omega,\tau),
\ea
where $\beta$ and $\phi$ are analytical in $\omega$ and assumed 
to be small. Eqs.\ (\ref{eqdyn}) and (\ref{abc1}) are expanded 
to first order in $\beta$ and $\phi$ about the 
uniformly translating circle and read
\ba
\label{An1}
{\rm Re}\left[\omega\;\partial_\tau\beta-\omega\partial_\omega\beta\right]
={\rm Re}\left[-\omega\partial_\omega\phi\right] 
~~~\mbox{for }\omega=e^{i\alpha},
\\
\label{An2}
\frac{\epsilon}{2}\;\left(\omega+\frac1{\omega}\right)
{\rm Re}\left[\omega^2\partial_\omega\beta\right]
={\rm Re}\left[\epsilon\;\omega\partial_\omega\phi+\phi\right].
\ea
By construction, $F(\omega)=\partial_\tau\beta-\partial_\omega\beta
+\partial_\omega\phi$ is analytical for $|\omega|<1$,
and Eq.\ (\ref{An1}) shows that ${\rm Re}[\omega\;F(\omega)]=0$
for $|\omega|=1$. Furthermore, it is clear that $\omega\;F(\omega)$
vanishes for $\omega=0$. Therefore, 
\be
\label{An3}
0 = \omega\;F(\omega) = 
\omega\;\left(\partial_\tau\beta-\partial_\omega\beta
+\partial_\omega\phi\right)
\ee
is valid on the whole closed unit disk. The corresponding analysis 
of Eq.\ (\ref{An2}) yields
\be
\label{An4}
\frac{\epsilon}{2}\;\left(\omega+\frac1{\omega}\right)\;
\omega^2\partial_\omega\beta=\epsilon\omega\partial_\omega\phi
+\phi+{\rm const.}
\ee
To this equation the operator $\omega\partial_\omega$ is applied,
and Eq.\ (\ref{An3}) is used to eliminate terms containing 
$\omega\partial_\omega\phi$. As a result, we find an equation only
for the function $\beta(\omega,\tau)$:
\ba
\label{Le0}
{\cal L}_\epsilon\;\beta&=&0,
\\
\label{Le1}
{\cal L}_\epsilon&=&
-\epsilon\;(1-\omega^2)\;\omega\;\partial_\omega^2
-(2+\epsilon-3\epsilon\;\omega^2)\;\partial_\omega
\nonumber\\
&&\qquad\qquad
+2\epsilon\;\omega\partial_\omega\;\partial_\tau
+2(1+\epsilon)\;\partial_\tau.
\ea
Eq.\ (\ref{Le0}) has to be solved for arbitrary initial conditions 
$\beta(\omega,0)$ that are analytical in some neighborhood of the unit disk.
The operator ${\cal L}_\epsilon$ conserves analyticity in time.
$\epsilon$ is a singular perturbation that multiplies 
the highest derivatives $\partial_\omega^2$ and 
$\partial_\omega\partial_\tau$. 
%

{\em The case $\epsilon=0$
is almost trivial, since ${\cal L}_\epsilon$ reduces to
\be
{\cal L}_0=2\;(\partial_\tau-\partial_\omega).
\ee }
Thus all solutions can be written as
\be
\beta(\omega,\tau)=\hat\beta(\omega+\tau),
\ee
where $\hat\beta(\zeta)$ is any function analytic in a neighborhood of
the unit disk $|\zeta|\le1$. The time evolution just amounts to a translation
along the strip $-1\le{\rm Re}~\zeta\le\infty$, $|{\rm Im}~\zeta|\le1$. 
Any singularity
of $\hat\beta$ at some finite point $\zeta$ on the strip will lead to
a breakdown of perturbation theory within finite time; this is the
generic behavior as found previously in the full nonlinear analysis
of this unregularized problem. Of course, there also exist
solutions that stay bounded for all times.

A different perspective on $\epsilon=0$ is that 
the Richardson moments are an infinite 
sequence of conserved quantities \cite{entov}. 
A reflection of this property
is that any polynomial $\beta(\omega,\tau)=\sum_{k=0}^N 
b_k(\tau)\omega^k$ for any $N$ with an appropriate choice of the 
time dependent functions $b_k(\tau)$ is an exact solution
for all times $\tau>0$, for linear perturbation 
theory (\ref{Le0}) as well as for the full nonlinear problem \cite{Bern1},
i.e., any truncation of the Laurent series (\ref{ft}) 
leads to exact solutions.

This observation suggests that an expansion in powers of $\omega$
is a natural ansatz also for nonvanishing (but small) $\epsilon$. 
Taking as initial condition some polynomial of order $N$,
one finds from the form (\ref{Le1}) of ${\cal L}_\epsilon$ that higher modes
$\omega^k$, $k>N$ are generated dynamically --- similarly to the 
daughter singularities in regularized viscous fingers \cite{Tanveer}.
When the expansion in $\omega$ is truncated at some arbitrary $N'$, 
it can be shown
that the problem for any truncation $N'$ and for any value $\epsilon>0$
has purely imaginary temporal eigenvalues. One would therefore expect
all eigensolutions for $\epsilon>0$ to be purely oscillating in time.
However, this behavior seems inconsistent with our exact solution 
for $\epsilon=1$. 

{\em For $\epsilon=1$ it turns out that the operator factorizes
\ba
\label{L1}
{\cal L}_1
&=&\Big[2\partial_\tau-(1-\omega^2)\partial_\omega\Big]\;
\Big[2+\omega\partial_\omega\Big].
\ea
which allows us to construct the general solution.}
We introduce the function
\be
\label{L2}
g(\omega,\tau)=[2+\omega\partial_\omega]\;\beta(\omega,\tau),
\ee
that obeys the equation
\be
\label{L3}
[2\partial_\tau-(1-\omega^2)\partial_\omega]\;g(\omega,\tau)=0.
\ee
The general solution of this equation reads
\be
\label{L4}
g(\omega,\tau)=G\left(\frac{\omega+T}{1+T\omega}\right),
~~~T=\tanh\frac{\tau}{2}.
\ee
The function $G$ is derived from the initial condition 
as
\be
G(\omega)=g(\omega,0)=[2+\omega\partial_\omega]\;\beta(\omega,0);
\ee
hence it is analytical in a neighborhood of the unit disk.
Finally, Eq.\ (\ref{L2}) is solved by
\be
\label{int2}
\beta(\omega,\tau)=\int_0^\omega \frac{x\;dx}{\omega^2}\;
G\left(\frac{x+T}{1+Tx}\right).
\ee

\begin{figure}

\includegraphics[width=8cm]{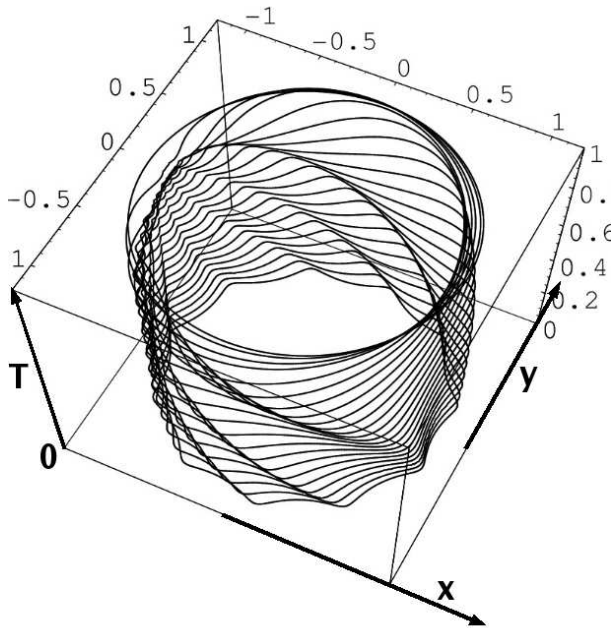}

\caption{Temporal evolution of a perturbed circle 
$f_t(\omega)-\tau=1/\omega+\beta(\omega,\tau)$ 
moving in the positive $x$-direction, according to
Eqs.~(\ref{Cir2}) and (\ref{L4})--(\ref{int2}).
The constant motion in time $\tau$ is subtracted.
The initial perturbation is a Fourier mode 
$\beta(\omega,0)=-0.5\;\omega^k/(k+2)$ with $k=10$.
The evolution during times $0\le\tau\le\infty$ corresponding to 
$0\le T\le 1$ is shown for time steps $T=$ 0, 0.05, 0.1, $\ldots$, 
0.85, 0.9, 0.95, 0.97, 0.99, 1.0.
}

\end{figure}

Now the one parameter family of mappings
\be
\omega \longrightarrow \zeta_T(\omega)=\frac{\omega+T}{1+T\omega},
~~~-1<T<1,
\ee
forms a subgroup of the automorphisms of the unit disk. Thus 
on the level of $G(\zeta)$, the dynamics amounts to a conformal 
mapping of the unit disk $|\omega|\le1$ onto itself.
This dynamics is somewhat distorted by the additional integration
(\ref{int2}) leading to $\beta(\omega,\tau)$, but it is easily
seen that $\beta(\omega,\tau)$ and $\partial_\omega\beta(\omega,\tau)$
are bounded uniformly in $\tau$ for $|\omega|\le 1$.
Hence, contrary to the unregularized problem for $\epsilon=0$,
only perturbations contribute that are bounded for all times.
Hence an infinitesimal perturbation can never form cusps.
Furthermore, the mapping $\omega\to\zeta_T(\omega)$ has fixed points
$\omega=\pm1$; and for $\tau\to\infty$, i.e., $T\to1$, it degenerates
to $\zeta_1(\omega)\equiv 1$, provided $\omega\ne -1$. 
We thus find the asymptotic behavior
\be
\beta(\omega,\tau)\stackrel{\tau\to\infty}{\longrightarrow}\frac{G(1)}{2},
\ee
independent of $\omega$ for any initial condition. 
Therefore asymptotically, the perturbation
just shifts the basic circular solution without change of shape.
Indeed, it is easily checked that any pronounced structure of
the initial perturbation that is not located right at the top at
$\omega=1$, is convected with increasing time toward $\omega=-1$
where it vanishes. This is an outflow of the simple dynamics 
of $G(\zeta)$ as pointed out above. Fig.~1 illustrates this behavior.

To summarize, we have found that the boundary condition (\ref{bcn})
at least for $\epsilon=1$ regularizes our problem in the sense
that an infinitesimal perturbation of a uniformly translating circle 
stays infinitesimal for all times and vanishes asymptotically 
for $\tau\to\infty$
up to an infinitesimal shift of the complete circle.
This statement is based on an exact analytical solution
for an arbitrary initial perturbation.
At the present stage, we have indications that this behavior of infinitesimal
perturbations is generic for $\epsilon>0$, while the solution is
unstable for $\epsilon=0$.
Furthermore, we expect that the convection of perturbations to
the back of the structure applies similarly for other shapes like
fingers. When applying the present calculation to streamers,
we in fact have to assume this to be true, since streamers are
typically not closed bodies, but rather the tips of ionized channels.
Finally, the behavior of finite perturbations and their nonlinear
analysis will require future investigations.


%

\newpage

%
%
%

\end{document}